\pdfoutput=1

\documentclass[aps,prb,reprint,showpacs,superscriptaddress,floatfix]{revtex4-1}

\usepackage{graphicx}
\usepackage{graphics}
\usepackage{amsmath}
\usepackage{amssymb}
\usepackage{amsfonts}
\usepackage{dcolumn}
\usepackage{latexsym}
\usepackage{rotating}
\usepackage{color}
\usepackage{latexsym}
\usepackage{bbm}
\usepackage{subfigure}
\usepackage{float}
\usepackage{epsfig}
\usepackage{epsf}
\usepackage{psfrag}
\usepackage{bm}
\usepackage{amsthm}
\usepackage{eucal}
\usepackage{mathrsfs}
\usepackage{url}
\usepackage{braket}
\usepackage{array}
\usepackage[utf8]{inputenc}
\usepackage{microtype}
\usepackage{multirow}

\usepackage{color} 

\usepackage{hyperref}
\hypersetup{
colorlinks=true,final=true,
        linkcolor=blue,
        citecolor=blue,
        filecolor=blue,
        urlcolor=blue,
}

\begin{document}
\title{Tuning the electronic and magnetic properties of NiBr$_2$ via pressure} 
\author{Soumen Bag}
\email{skbag@asu.edu}
\affiliation{Department of Physics, Arizona State University, Tempe, AZ - 85287, USA}

\author{Jesse Kapeghian}
\affiliation{Department of Physics, Arizona State University, Tempe, AZ - 85287, USA}

\author{Onur Erten}
\email{onur.erten@asu.edu}
\affiliation{Department of Physics, Arizona State University, Tempe, AZ - 85287, USA}

\author{Antia S. Botana}
\email{antia.botana@asu.edu}
\affiliation{Department of Physics, Arizona State University, Tempe, AZ - 85287, USA}

\date{\today}

\begin{abstract}
Transition metal dihalides (MX$_2$, M= transition metal, X= halide) have attracted much attention recently due to their intriguing low-dimensional magnetic properties. Particular focus has been placed in this family in the context of multiferroicity-- a common occurrence in MX$_2$ compounds
that adopt non-collinear magnetic structures. One example of helimagnetic multiferroic material in the dihalide family is represented by NiBr$_2$. Here, we study the evolution of the electronic structure and magnetic properties of this material under pressure using first-principles calculations combined with Monte Carlo simulations. Our results indicate there is significant magnetic frustration in NiBr$_2$ due to the competing interactions arising from its underlying triangular lattice. 
This magnetic frustration increases with pressure and is at the origin of the helimagnetic order. Further, pressure causes a sizable increase in the interlayer interactions. Our Monte Carlo simulations show that a large (3-fold) increase in the helimagnetic transition temperature can be achieved at pressures of around 15 GPa. This indicates that hydrostatic pressure can indeed be used as a tuning knob to increase the magnetic transition temperature of NiBr$_2$.

\end{abstract}
\maketitle
\section{Introduction}\label{introduction} 

 Two-dimensional (2D) van der Waals (vdW) magnets have been intensively studied as they provide powerful platforms to explore novel physical phenomena and to implement intriguing applications \cite{Blei_APR_2021, Gibertini2019}. Transition metal dihalides represent an emerging class of 2D vdW magnets that can exhibit multiferroic order and non-collinear spin textures \cite{mcguire2017crystal}. Within this family, the magnetic semiconductor NiI$_2$ has been the subject of much research in recent years\cite{Amoroso_NC_2020,  Ju2021, Song_Nat_2022, Fumega_2022, lebedevnii2, Das_PRB2024}. In the bulk, two magnetic phase transitions take place in this material: one at 75 K to a collinear antiferromagnetic (AFM) state and one at 60 K to a helimagnetic state \cite{Billerey_PLA_1977,BILLEREY198059}. This noncollinear magnetic state simultaneously hosts a spin-induced ferroelectric polarization tunable with magnetic field, making NiI$_2$ a type-II multiferroic \cite{ Kurumaji_PRB_2013}. Recently, it has been demonstrated that the  multiferroic phase in NiI$_2$ persists from the bulk to the single-layer limit \cite{Song_Nat_2022}. Further, it has been shown that a significant enhancement of the helimagnetic order in bulk NiI$_2$ can be achieved with hydrostatic pressure \cite{Occhialini_arXiv_2023, kapeghian2023effects} all the way up to  132 K at 5 GPa.  

\begin{figure}
\centering
\includegraphics[width=\columnwidth]{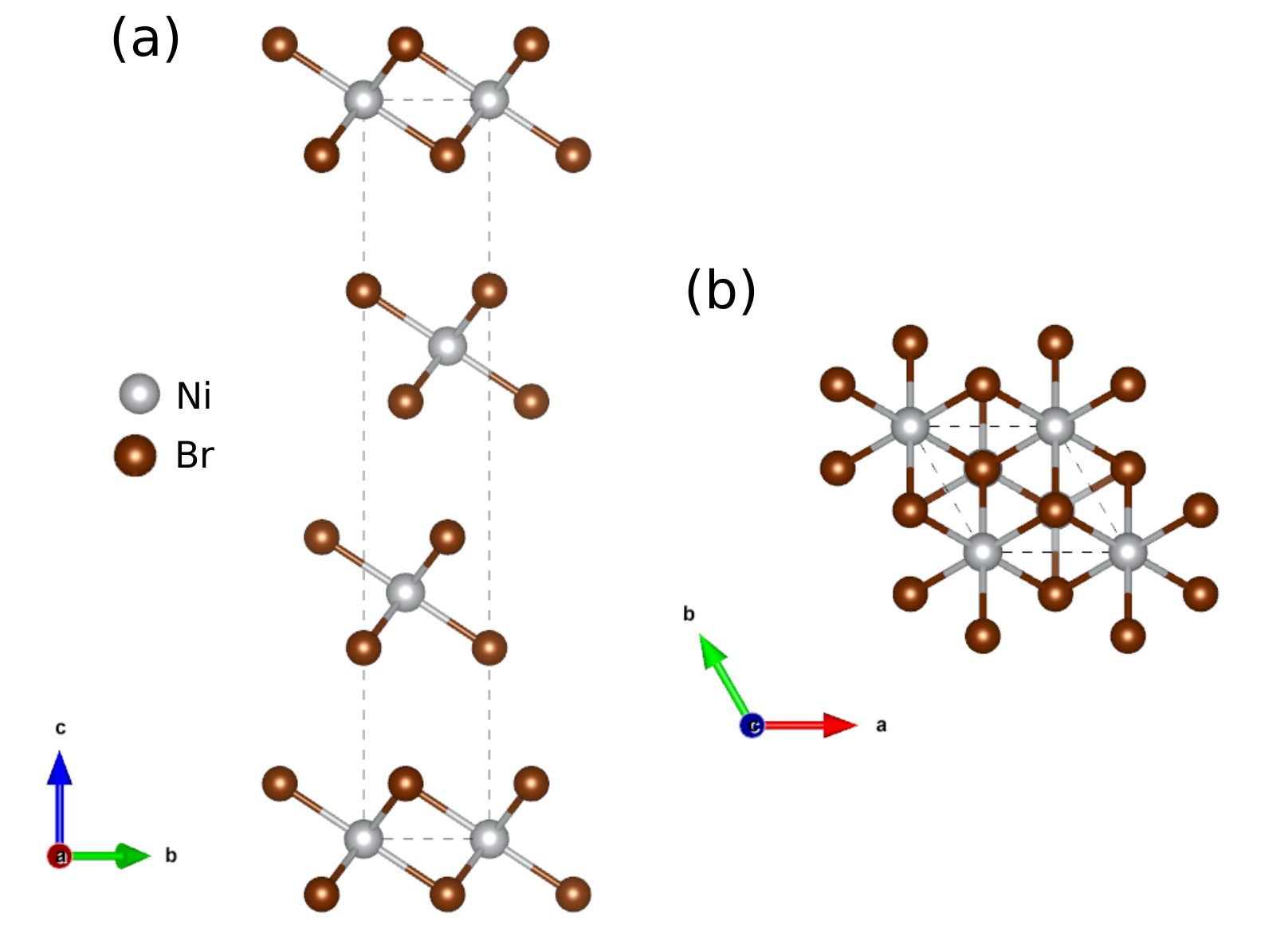}
\caption{Crystal structure of bulk NiBr$_{2}$.  Out-of-plane (a) and in-plane (b) views of the $R \overline{3} m$ structure. Ni atoms are represented by gray spheres, while brown spheres represent Br atoms. A black dotted line marks the unit cell boundaries.}
\label{fig:1}
\end{figure}
 
Given that 2D multiferroics would provide disruptive
possibilities to electrically control magnetic order, it is interesting to further explore other candidate materials in this context. An obvious choice is NiBr$_2$, a related compound from the dihalide family. NiBr$_2$ crystallizes in a CdCl$_2$ structure (space group $R \overline{3} m$)\cite{P_Day_1976, Nasser_SSC_1992}
as depicted in Fig. \ref{fig:1}. Its structure is formed by edge-sharing NiBr$_6$ octahedra (forming a triangular lattice) that stack along the $c$ axis with weak vdW
bonding. The Ni$^{2+}$ (S=1) Ni ions order antiferromagnetically
at T$_{N,1}$ = 52 K \cite{P_Day_1976, P_Day_1980, ADAM19801}. As in NiI$_2$, this collinear AFM phase consists of ferromagnetic planes coupled antiferromagnetically out of plane. At T$_{N, 2}$= 23 K, a second transition occurs to a spin-spiral order\cite{P_Day_1980, ADAM19801}. 
Interestingly, akin to NiI$_2$, NiBr$_2$ also develops a ferroelectric polarization in its helimagnetic low-temperature ground state \cite{tokunaga2011multiferroicity}. 
Notably, the helimagnetic transition temperature of NiBr$_2$ is considerably lower than that of NiI$_2$ but hydrostatic pressure could in principle be exploited as a means to enhance it.  

Here, we study the effects of hydrostatic pressure on the magnetic properties of bulk NiBr$_2$ using a combination of first-principles calculations and Monte Carlo simulations. 
Our results indicate that there is a substantial magnetic frustration in NiBr$_2$ (that increases with pressure) arising from the competition between the intralayer ferromagnetic nearest-neighbor interaction ($J^{\parallel 1}$) and the antiferromagnetic third nearest-neighbor interaction ($J^{\parallel 3}$). Such magnetic frustration is at the origin of its helimagnetic ground state whose transition temperature we can accurately reproduce at ambient pressure using Monte Carlo simulations. We find that pressure has a significant effect on the interlayer coupling ($J^{\perp \mathrm{eff}}$), but also on some of the leading intralayer interactions. 
Using the first-principles-derived magnetic constants, Monte Carlo simulations reveal a 3-fold increase in the helimagnetic transition temperature of NiBr$_2$ at a modest pressure of 15 GPa.

\section{Computational Methods} \label{CM}

\textit{First-Principles Calculations.} We conducted density functional theory (DFT)-based calculations in NiBr$_2$ using the projector augmented wave (PAW) method \cite{Kresse_PRB_1999} as implemented in the VASP code \cite{Kresse_PRB_1996,Kresse_CMS_1996}. The wave functions were expanded in the plane-wave basis with a kinetic-energy cut-off of 500 eV. We considered the $3p$, $3d$, and $4s$ orbitals ($3p^{6} 3d^{8} 4s^{2}$ configuration) as valence states for the Ni atoms. Meanwhile, for the Br atoms, we considered the $4s$ and $4p$ orbitals ($4s^{2} 4p^{5}$ configuration) as valence states. 

 Hydrostatic pressure was applied in 5 GPa increments (up to 15 GPa), conducting full structural relaxations. The optimization of the bulk unit cells at each pressure involved optimizing atomic positions, cell shape, and cell volume, but focusing exclusively on the rhombohedral phase. The energy and force minimization tolerances were set at $10^{-10}$ eV and $10^{-3}$ eV/\AA, respectively. The calculations were done using the Perdew-Burke-Ernzerhof (PBE) \cite{Perdew_PRL_1996} version of the generalized gradient approximation (GGA) functional, with the inclusion of the DFT-D3 van der Waals correction \cite{Grimme_JCP_2010}. Additionally, we incorporated an on-site Coulomb repulsion parameter ($U$) using the Liechtenstein \cite{Liechtenstein_PRB_1995} approach to account for correlation effects in the Ni-$d$ electrons \cite{Rohrbach_JPCM_2003}. The $U$ and Hund's coupling $J_{\mathrm{H}}$ values utilized in all the calculations presented in the main text ($U=3.9$ eV and $J_{\mathrm{H}}=0.79$ eV) were derived from constrained random phase approximation (cRPA) calculations \cite{Riedl_PRB_2022}. For all of the relaxations, we fixed the magnetic configuration to an AFM state comprised of FM planes coupled AFM out of plane. To accommodate the AFM ordering, we employed a $1\times1\times2$ supercell and conducted Brillouin zone (BZ) sampling using a $40\times40\times4$ Monkhorst-Pack k-mesh centered on the $\Gamma$ point. This AFM order aligns with the $c$-component of the magnetic propagation vector ($\sim$ 3/2) \cite{mcguire2017crystal}. 

Finally, we computed the exchange couplings and anisotropies for NiBr$_{2}$ using the four-state method, extensively detailed in Refs. [\onlinecite{Xiang_DT_2013,Xiang_PRB_2011,Sabani_PRB_2020,Xu_NPJCM_2018,Xu_PRB_2020}]. This method relies on performing total energy mappings through noncollinear magnetic DFT calculations with spin-orbit coupling (SOC). Each magnetic interaction parameter is associated with the energies of four distinct magnetic configurations, wherein the directions of the magnetic moments are constrained, and large supercells are employed to prevent coupling between distant neighbors. Using this methodology, intralayer (interlayer) magnetic constants were calculated for each pressure.

\textit{Monte Carlo Simulations.} We employed the Matjes\cite{matjes} code to conduct Monte Carlo simulations in NiBr$_2$ to further investigate its magnetic response with pressure. Around $\sim 10^6$ thermalization steps were executed at each temperature, followed by $\sim 10^4$ Monte Carlo steps for statistical averaging. The simulations utilized a standard Metropolis algorithm on supercells with dimensions $L \times L \times 4$ and periodic boundary conditions. To determine the supercell size $L$, we adopted the criterion $L \simeq nL_{\mathrm{m.u.c.}}$, where $n$ is an integer, and $L_{\mathrm{m.u.c.}}$ represents the minimum lateral size of the magnetic unit cell. 
The length of the magnetic unit cell $L_{\mathrm{m.u.c.}}$ was estimated as $L_{\mathrm{m.u.c.}} \sim 1 /q^{\parallel}$, where $q^{\parallel}$ denotes the magnitude of the in-plane component of the magnetic propagation vector derived as $q^{\parallel} = \frac{1}{2\pi}\arccos{\left[ \left( 1 + \sqrt{1 - 2 (J^{\parallel 1}/J^{\parallel 3}) } \right) / 4 \right]}$\cite{Hayami_PRB_2016,Batista_RPP_2016}.


\section{Results}\label{Result}

We start by introducing the microscopic model that we will follow to obtain the relevant magnetic interactions for NiBr$_{2}$, given by the following Heisenberg Hamiltonian between localized spins
\textbf{S$_{i}$} that we split into intra- and interlayer contributions expressed as $H^{\parallel}$ and $H^{\perp}$, respectively,

\begin{equation} \label{eq:1}
H^{\parallel} = \frac{1}{2} \sum_{i \neq j} \mathbf{S}_{i} \cdot \mathbf{J}^{\parallel}_{ij} \cdot \mathbf{S}_{j} + \sum_{i} \mathbf{S}_{i} \cdot \mathbf{A}_{i} \cdot \mathbf{S}_{i},
\end{equation}

\begin{equation} \label{eq:2}
H^{\perp} = \frac{1}{2} \sum_{i,j} J^{\perp}_{ij}  \mathbf{S}_{i} \cdot \mathbf{S}_{j},
\end{equation}
Here, the indices $i$ and $j$ refer to the Ni atom sites. In Eq. \ref{eq:1},  \textbf{A$_i$} denotes the on-site or single-ion anisotropy (SIA) and \textbf{J$_{ij}$} represents the intralayer exchange coupling interaction tensor. The latter can be decomposed into two contributions for NiBr$_2$: an isotropic coupling term, and an anisotropic symmetric term (the antisymmetric term which corresponds to the Dzyaloshinskii–Moriya interaction vanishes in NiBr$_2$ due to the presence of inversion symmetry). $J^{\perp}_{ij}$ in Eq. \ref{eq:2} represents the isotropic interlayer exchange constant between spins $\mathbf{S}_{i,j}$. We consider up to third nearest-neighbor isotropic exchanges both in-plane and out-of-plane. The full tensor is only taken into account for the in-plane nearest-neighbor exchange interaction. The factors of 1/2 are used to account for double-counting. The sign conventions used here are as follows: a positive (negative) isotropic exchange interaction favors an antiparallel (parallel) alignment of spins and a positive (negative) scalar single-ion parameter indicates an easy-plane (easy-axis) anisotropy.

\begin{figure}[htbp!]
\begin{center}
\includegraphics[width=1.0\linewidth]{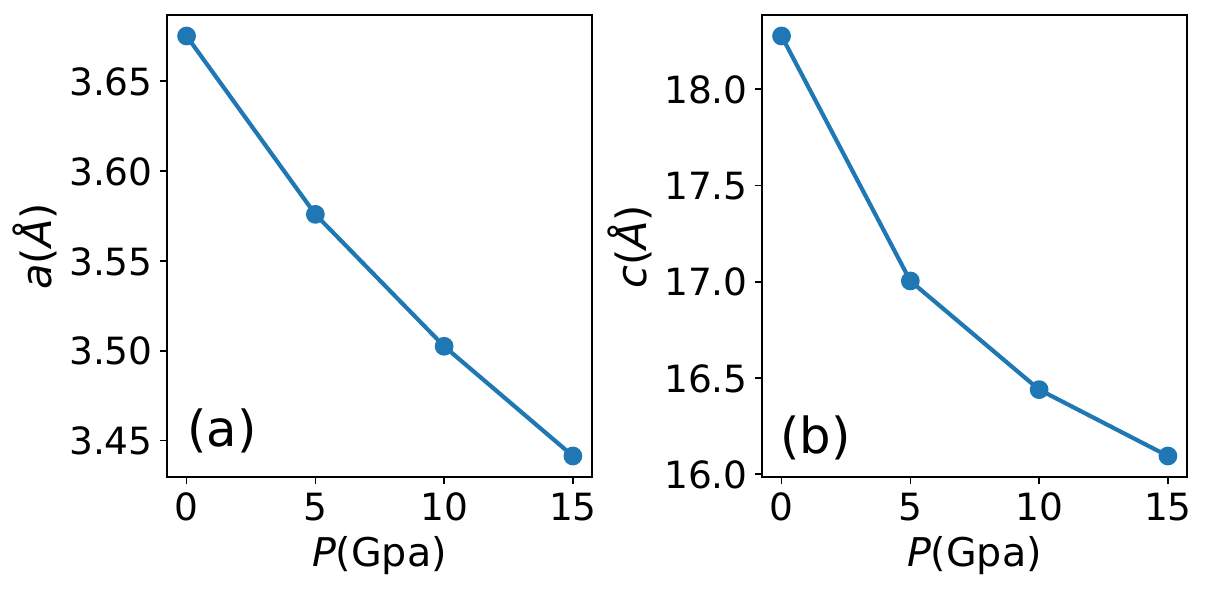}
\end{center}
\label{DHM11}
\caption{First-principles derived (a) in-plane  and (b) out-of-plane relaxed lattice parameters as a function of pressure (P) for bulk NiBr$_2$.}
\label{fig:2}
\end{figure}




Before moving into the evolution of the relevant magnetic parameters, we start by describing the evolution of the structural properties of NiBr$_2$ under pressure. Figure \ref{fig:2} displays the relaxed lattice parameters of NiBr$_2$ as a function of pressure obtained from our first-principles calculations using the computational parameters described in Section \ref{CM}. 
Upon applying hydrostatic pressure, both the in-plane (see Fig. \ref{fig:2}(a)) and out-of-plane (see Fig. \ref{fig:2}(b))  lattice parameters decrease monotonically, with a much larger decrease in the out-of-plane lattice parameter, as expected for a van der Waals material. Specifically, $a$ decreases from 3.67 \AA~ at ambient pressure to 3.44 \AA~ at 15 GPa while $c$ decreases from 18.27 \AA ~to 16.09 \AA ~at 15 GPa.

The basic evolution of the electronic structure is shown in Appendix \ref{Appband}. Up to the highest pressures studied here, NiBr$_2$ remains insulating within our GGA+$U$ calculations (the gap can only be closed at $\sim$ 80 GPa). 
The derived magnetic moment for the Ni atoms is $\sim 1.54~ \mu_{\mathrm{B}}$ at all pressures, consistent with high-spin Ni$^{2+}$ but with a slightly reduced value with respect to the nominal one due to hybridization with the Br ligands (with moments $\sim 0.17 \, \mu_{\mathrm{B}}$ at all pressures). Our first-principles derived magnetic moments are in good agreement with the ordered experimental Ni moment values obtained at ambient pressure  $\sim 1.5 \, \mu_{\mathrm{B}}$  \cite{NiBr2Film, mcguire2017crystal}.

\renewcommand{\arraystretch}{1.3}
\begin{table}
\centering
\begin{tabular}{m{3.0em} m{3.75em} m{3.75em} m{3.75em} c}
\hline
\hline
\multicolumn{5}{c}{Isotropic intralayer exchanges} \\
$P$ & $J^{\parallel 1}$ & $J^{\parallel 2}$ & $J^{\parallel 3}$ & $J^{\parallel 3}/J^{\parallel 1}$ \\
\hline
0   & -3.19 & -0.05 & 1.56 & -0.49 \\ 
5   & -3.71       & -0.06      & 2.25      & -0.61 \\
10  & -4.24       & -0.14      & 2.92      & -0.69 \\
15  & -4.68       & -0.17      & 3.74      & -0.8\\
\end{tabular}
\begin{tabular}{lcccccc}
\hline
\multicolumn{7}{c}{SIA and intralayer TSA} \\
$P$ & $A$ & $J^{\mathrm{S} \parallel 1}_{xx}$ & $J^{\mathrm{S} \parallel 1}_{yy}$ & $J^{\mathrm{S} \parallel 1}_{zz}$ & $J^{\mathrm{S} \parallel 1}_{yz}$ & $J^{\mathrm{S} \parallel 1}_{yz}/J^{\parallel 1}$ \\
\hline
0  & 0.0 & -0.04 & 0.04 & 0.0 & -0.06 & 0.019\\ 
5  & 0.0      & -0.05 & 0.04 & 0.0 & -0.07 & 0.018 \\ 
10 & 0.0      & -0.05 & 0.05 & 0.0 & -0.07 & 0.018\\ 
15 & 0.03     & -0.2 & 0.13 & 0.07& -0.08 & 0.017 \\ 
\hline
\end{tabular}
\begin{tabular}{m{2.78em} m{2.78em} m{2.78em} m{2.78em} m{2.78em} c}
\multicolumn{6}{c}{Isotropic interlayer exchanges} \\
$P$ & $J^{\perp 1}$ & $J^{\perp 2}$ & $J^{\perp 3}$ & $J^{\perp \mathrm{eff}}$ & $J^{\perp 2}/J^{\parallel 1}$ \\
\hline
 0 & 0.01   & 0.62 & 0.17 & 0.96  & -0.19  \\ 
 5 & 0.01 & 1.7  & 0.42 & 2.54 & -0.46\\ 
10 & 0.02  & 2.84 & 0.67 & 4.19 & -0.67\\ 
15 & 0.12  & 4.02 & 0.9  & 5.94 &  -0.85\\ 
\hline
\hline
\end{tabular}
\caption{Calculated NiBr$_2$ isotropic intralayer exchange interactions (top panel), interlayer isotropic interactions (bottom panel), SIA (A), and first-nearest neighbor in-plane two-site anisotropy (TSA) constants (middle panel) in a cartesian $x, y, z$ reference system where $x$ was chosen to be parallel to the Ni-Ni bonding vector for different pressures ($P$). $J^{\perp eff}$ represents the effective interlayer exchange and is obtained as $J^{\perp 1} + J^{\perp 2} +2J^{\perp 3}$, where the coefficient in the last term arises because there are twice as many out-of-plane third nearest-neighbors as first- and second nearest-neighbors. Pressure is in units of GPa, and exchange constants are given in units of meV.}
\label{table:1}
\end{table}

\begin{figure}[H]
\includegraphics[width=0.9\columnwidth]{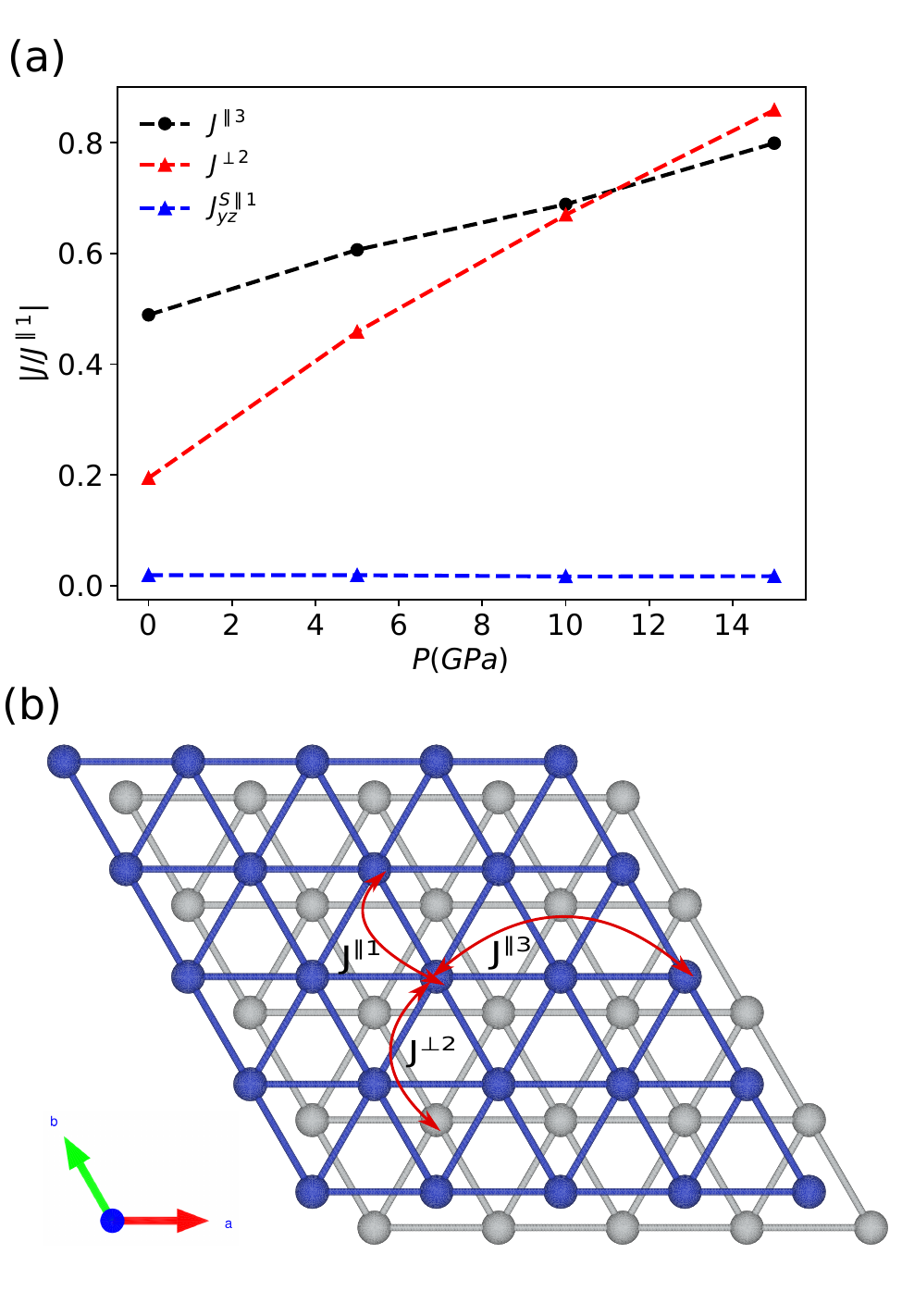}
\caption{(a) Pressure evolution of the three relevant ratios of exchange couplings $J^{\parallel 3}$/$J^{\parallel 1}$, $J^{\perp 2}$/$J^{\parallel 1}$,  $J^{S \parallel  1}_{yz}$/$J^{\parallel 1}$ for bulk NiBr$_{2}$. (b) The paths for the dominant exchange interactions $J^{\parallel 1}$, $J^{\parallel 3}$ and $J^{\perp 2}$ are shown over the triangular arrangement of Ni atoms in NiBr$_2$ (with gray spheres corresponding to Ni atoms in the bottom layer and blue spheres
to those in the top layer).}
\label{fig:4}
\end{figure}

After establishing the basics of the evolution of the structure and electronic structure in the AFM collinear phase, next we move on to the calculations of the magnetic coupling constants for NiBr$_{2}$ using the four-state method (we followed the implementation used for other dihalides as described in Refs. \onlinecite{Amoroso_NC_2020, Gorkan_PRB2023, kapeghian2023effects}). Table \ref{table:1} presents the computed intralayer and interlayer magnetic parameters introduced in Eqs. \ref{eq:1} and \ref{eq:2}, as well as relevant ratios between magnetic couplings for pressures up to 15 GPa. Fig \ref{fig:4}(a) shows the evolution of these relevant magnetic exchange ratios as a function of pressure while Fig \ref{fig:4}(b) shows the paths for the dominant exchange interactions. At ambient pressure, the largest exchange interaction in NiBr$_2$ is the ferromagnetic (FM) intralayer first nearest-neighbor exchange ($J^{\parallel 1}$ $\sim$ -3.2 meV). The second nearest-neighbor exchange is vanishingly small and FM, while the third nearest-neighbor intralayer exchange is  AFM and sizable ($J^{\parallel 3}$ $\sim$ 1.6 meV). The derivation of a large $J$ for third (vs. second) nearest-neighbors is consistent with previous results obtained for dihalide monolayers in Ref. \onlinecite{Riedl_PRB_2022}: for second nearest-neighbors, only the $t_{2g}$-$e_g$ hopping is relevant at $d^8$ filling, leading
to a weak FM interaction, while for third nearest-neighbors, there are large $e_g$-$e_g$ hoppings that arise from hopping paths that are ligand-assisted. Importantly, the competition between intralayer FM $J^{\parallel 1}$ and AFM $J^{\parallel 3}$ (measured by the ratio $J^{\parallel 3}/J^{\parallel 1} = -0.5$) results in a strong magnetic frustration which favors the realization of the non-collinear magnetic ground state of NiBr$_{2}$ \cite{Rastelli_PBC_1979}. 
Another critical parameter in the context of magnetic exchanges is the ratio $J^{\mathrm{S} \parallel 1}_{yz}/J^{\parallel 1}$ which gauges the canting of the two-site anisotropy axes from the direction perpendicular to the layers\cite{Amoroso_NC_2020}. This ratio is estimated to be very small in NiBr$_2$ $J^{\mathrm{S} \parallel 1}_{yz}/J^{\parallel 1} = 0.019$. Moving to the interlayer exchange interactions, they are all AFM in nature with the second nearest-neighbor $J^{\perp 2}$ being the dominant one $\sim$ 0.6 meV. If we look at the ratio between the dominant intra- vs. interlayer interactions $J^{\perp 2}/J^{\parallel 1}$$\sim$ - 0.2 at ambient pressure.

\begin{figure*}
\includegraphics[width=1.8\columnwidth]{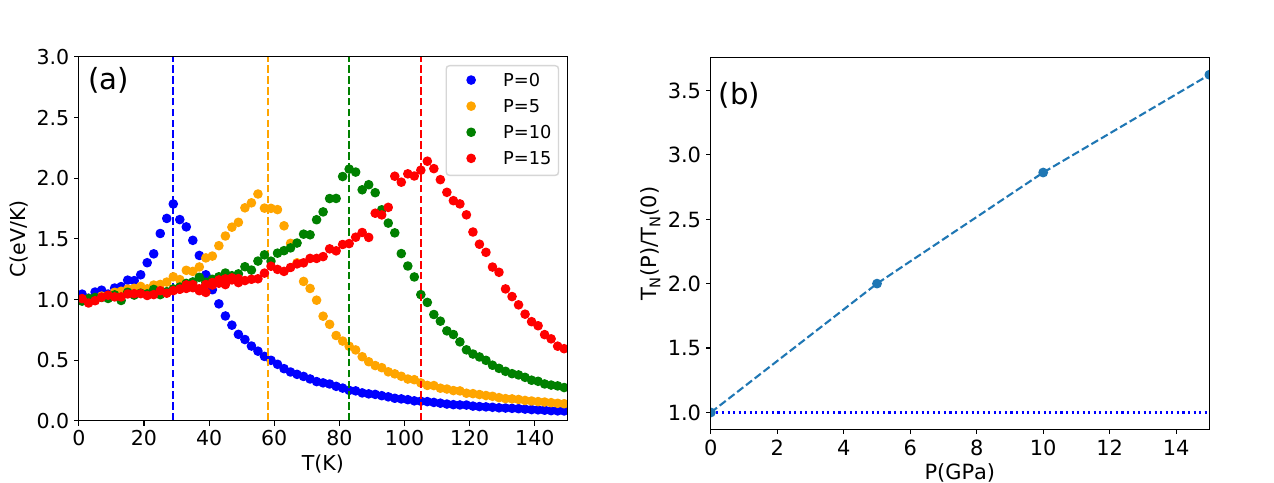}
\caption{(a) Specific heat $C$ of bulk NiBr$_{2}$ as a function of temperature $T$  for various pressures P = 0, 5, 10, and 15 GPa obtained from Monte Carlo simulations. The dashed lines indicate the critical temperature $T_{\mathrm{N}}$ at each pressure: 29, 58, 83, and 105 K for 0, 5, 10, and 15 GPa, respectively. (b) Ambient-pressure-normalized critical temperature values for bulk NiBr$_{2}$ as a function of pressure $P$.}
\label{fig:5}
\end{figure*}

The signs of the dominant intra- and interlayer interactions do not change with pressure but the magnitude of the isotropic magnetic constants increases considerably. For the first nearest-neighbor isotropic exchange $J^{\parallel 1}$$_{15 GPa}$= 1.5 $J^{\parallel 1}$$_{0 GPa}$, for the third nearest-neighbor isotropic exchange $J^{\parallel 3}$$_{15 GPa}$= 2.4 $J^{\parallel 3}$$_{0 GPa}$, while the second nearest-neighbor $J^{\parallel 2}$ remains vanishingly small at all pressures. The pressure-dependent response of the dominant intralayer isotropic exchanges can be understood by looking at the relevant hopping amplitudes, as we showed before for NiI$_2$ \cite{kapeghian2023effects}. For $J^{\parallel 1}$ there are two primary contributions, one being FM (mainly arising from the hopping process between $t_{2g}$ and $e_{g}$ states via the ligand $p$ states) and the other AFM (mainly arising from direct $d-d$ overlap between Ni $t_{2g}$-like states). With increasing pressure, the FM contribution increases at a faster rate resulting in an overall increase of $J^{\parallel 1}$ even though the competition due to the AFM hoppings still persists. In contrast, $J^{\parallel 3}$ exhibits solely AFM contributions originating from $e_{g}$-$e_g$ hoppings, as mentioned above, without FM contributions. In this manner, the $J^{\parallel 3}/J^{\parallel 1}$ ratio undergoes a sizable increase with pressure from -0.5 at ambient pressure to -0.8 at 15 GPa (see Fig. \ref{fig:4}(a)). The single-ion anisotropy is negligible at all pressures, and the intralayer anisotropic exchanges($J^{\mathrm{S} \parallel 1}_{yz}$) exhibit minimal changes with pressure as well. The ratio $J^{\mathrm{S} \parallel 1}_{yz}/J^{\parallel 1}$ remains nearly constant (and small) up to 15 GPa, as depicted in Fig. \ref{fig:4}(a).

Regarding the interlayer exchanges, the signs of the dominant interlayer isotropic exchange interactions persist as well: both $J^{\perp 2}$  and $J^{\perp 3}$ remain antiferromagnetic in the pressure range studied here, even though they increase sizably with pressure ($J^{\perp 1}$ remains small in comparison). This substantial increase is particularly noticeable for the dominant second nearest-neighbor interlayer exchange $J^{\perp 2}$$_{15 GPa}$= 6.5$J^{\perp 2}$$_{0 GPa}$. Such a large increase can be attributed to the significant decrease in the $c$ lattice parameter with pressure described above, as expected in a vdW material. Importantly, $J^{\perp 2}$ at 15 GPa becomes the second largest interaction overall, closely competing in value with $J^{\parallel 1}$ (see Fig. \ref{fig:4}(a)). In fact, if considering the overall effective interlayer exchange $J^{\perp \mathrm{eff}}$, it surpasses the dominant intralayer $J$  at $\sim$ 10 GPa as shown in Table \ref{table:1}.


The magnetic constants derived from the four-state method for NiBr$_2$ were subsequently used in Monte Carlo simulations. At low temperatures, we confirm that the derived magnetic ground state is a spin spiral (this is consistent with previous DFT-based studies that reported a spin-spiral ground state in monolayer NiBr$_2$\cite{amoroso2021interplay, sodequist2023type,DaWeiWu_PRB2023}, see the corresponding magnetic texture and structure factor in Appendix \ref{AppF}). From our pressure-dependent specific heat calculations we can clearly observe a magnetic transition at a temperature $T_{\mathrm{N}}$, indicated by the dashed vertical line in Fig. \ref{fig:5}(a), that increases monotonically with pressure. Although some double-peak
structure can be observed (that was also obtained in similar calculations for NiI$_2$\cite{kapeghian2023effects}) we focus here on a qualitative understanding of
the trends in the magnetic response with pressure, rather than pursuing a
quantitative description of the two magnetic transitions (more rigorous statistical measures would be needed
 to assign physical meaning to the aforementioned double-peak structure, which we leave for future work). Fig. \ref{fig:5}(b) clearly shows the monotonic increase of  $T_{\mathrm{N}}$ as a function of pressure, with the data points being normalized relative to the value calculated at ambient pressure  ($T_{\mathrm{N}}$(0 GPa)=29 K, very close to the experimentally derived value of $\sim$ 23 K). Notably, our calculated $T_{\mathrm{N}}$ undergoes a three-fold increase between 0 and 15 GPa (rising from 29 to 105 K). 

 Importantly, the increase in the $J^{\parallel 3}/J^{\parallel 1}$ ratio with pressure we have found in NiBr$_2$ has important implications for the  helimagnetic propagation
vector and likely for the related spin-induced ferroelectric polarization. As mentioned in Section \ref{CM}, the in-plane component of the magnetic propagation vector can be determined as $q ^{\parallel} = 2\arccos{\left[ \left( 1 + \sqrt{1 - 2 (J^{\parallel 1}/J^{\parallel 3}) } \right) / 4 \right]}$, the related spin-induced ferroelectric order can be estimated as P $\propto$ sin($q$) by the generalized Katsura-Nagaosa-Balatsky model \cite{kurumaji2013magnetoelectric, Song_Nat_2022, gNKB_xiang_2011}. The observed increase in
$J^{\parallel 3}/J^{\parallel 1}$  with pressure favors a larger $q$ (shorter in-plane spiral pitch) which, potentially, can then give rise to a larger spin-induced polarization (see Appendix \ref{AppD} for further details). 



\section{Summary}

To summarize, we employed first-principles calculations combined with Monte Carlo simulations to investigate the impact of hydrostatic pressure on the magnetic properties of bulk NiBr$_{2}$.  Using the four-state method, we computed the intralayer and interlayer exchange parameters (up to third nearest neighbors) of the low energy effective spin model for bulk NiBr$_{2}$. The low-temperature magnetic ordering corresponds to a spin spiral that is governed by the magnetic frustration between the two dominant in-plane exchange terms ($J^{\parallel 1}$ and $J^{\parallel 3}$), exhibiting different signs (ferro- and antiferromagnetic, respectively). The interlayer exchanges were identified as antiferromagnetic, with $J^{\perp 2}$ being the dominant interaction. With increasing pressure, all the dominating exchange couplings ($J^{\parallel 1}$, $J^{\parallel 3}$ and $J^{\perp 2}$) increase monotonically, and consequently, the (heli)-magnetic ordering temperature increases. These results suggest that hydrostatic pressure holds promise as a means to enhance the magnetic response of NiBr$_{2}$. Even though we do not analyze here the corresponding induced electric polarization, we anticipate that pressure could also potentially enhance the concomitant multiferroic response of NiBr$_{2}$. 


\section{ACKNOWLEDGMENTS}
SB, JK, OE and AB acknowledge support from NSF Grant No. DMR-2206987 and the ASU Research Computing Center for high-performance computing resources.


\bibliography{halaide.bib}
\clearpage
\newpage
\onecolumngrid 

\appendix

\section{ Band Structure evolution with pressure for NiBr$_{2}$}  \label{Appband}

\begin{figure}
\includegraphics[width=\columnwidth]{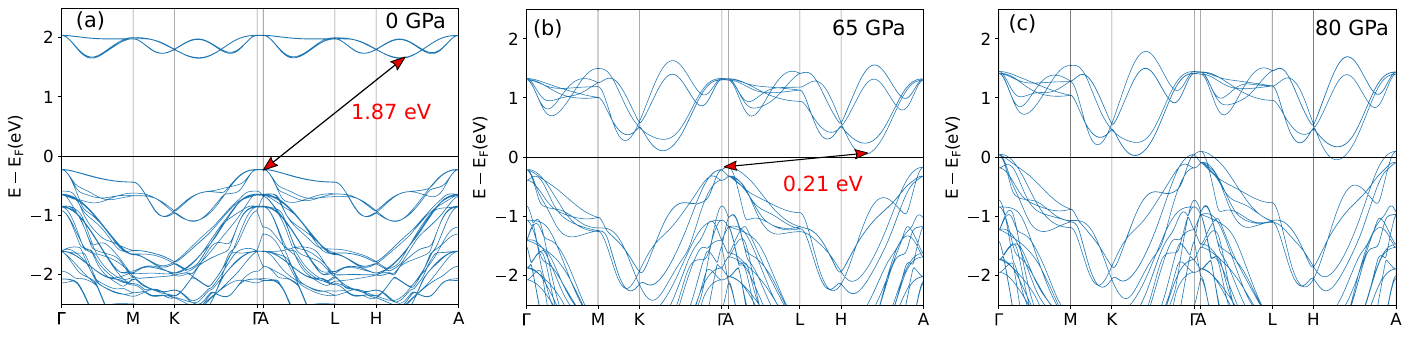}
\caption{GGA-PBE+D3+$U$ band structure plots for bulk NiBr$_{2}$ (calculated with AFM order and $U=3.9$ eV) at ambient pressure (a), 65 GPa (b) and 80 GPa (c) where the energy band gaps are indicated in the insulating 0 and 65 GPa cases. Reciprocal space coordinates: $\Gamma=(0,0,0)$, $\mathrm{M}=(1/2,0,0)$, $\mathrm{K}=(1/3,1/3,0)$, $\mathrm{A}=(0,0,1/2)$, $\mathrm{L}=(1/2,0,1/2)$, $\mathrm{H}=(1/3,1/3,1/2)$}
\label{fig1:app}
\end{figure}

Fig \ref{fig1:app} shows the evolution of the band structure along high-symmetry directions for  NiBr$_{2}$ in the collinear an AFM state (consisting of ferromagnetic planes coupled antiferromagnetically out-of-plane) under hydrostatic pressure. The band gap can only be closed at $\sim$ 80 GPa. 

Table \ref{table:4} contains the corresponding indirect band gap $E_{gap}$ of bulk NiBr$_{2}$ as function of pressure. 

\renewcommand{\arraystretch}{1.3}
\begin{table}[H]
\centering
\begin{tabular}{lccc}
\hline
\hline
$P$ (GPa) & E$_{\rm gap}$ \\
\hline
0  & 1.8737  \\
5  & 1.7282  \\
10 & 1.5999 \\
15 & 1.4630 \\
40 & 0.8075 \\
65 & 0.2137 \\
80 & 0.0  \\
\hline
\hline
\end{tabular}
\caption{Indirect band gap $E_{gap}$ of bulk NiBr$_{2}$ as function of pressure. Here, a zero band gap value corresponds to a metallic state.}
\label{table:4}
\end{table}

\section{Magnetization textures of monolayer NiBr$_{2}$} \label{AppF}

\begin{figure}[H]
\includegraphics[width=1.0\textwidth]{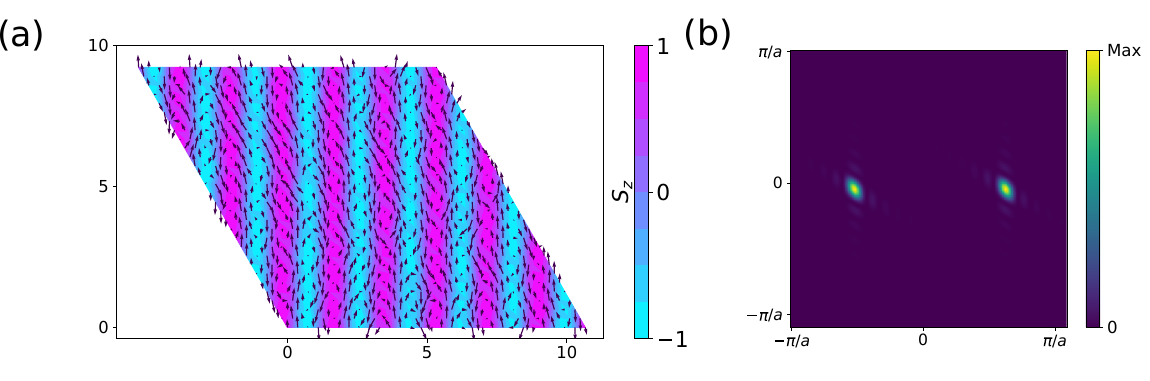}
\caption{ (a) Magnetization texture: black arrows represent the in-plane {s$_x$, s$_y$} spin components; the colormap indicates the
out-of-plane s$_z$ spin component and (b) Spin structure factor of monolayer NiBr$_2$ at ambient pressure at  1 K.}
\label{fig:E1}
\end{figure}

Fig \ref{fig:E1}(a) shows the magnetization texture of monolayer NiBr$_2$ ($10\times10$ supercell) at P=0 and T=1 K, which exhibits a spin-spiral structure along the $x$ direction. The spin-spiral structure along the $x$ direction is confirmed by the spin structure factor data shown in Fig \ref{fig:E1}(b). The spin structure factor for momentum $\textbf{q}$ is defined as
\begin{equation}
    S(\textbf{q}) = \frac{1}{N} \sum_{\alpha=x,y,z} \langle | \sum_i s_{i\alpha} e^{-i\textbf{q}\cdot\textbf{r}_i}|^2 \rangle,
\end{equation}
where N = $L^2$ is the total number of spins and $s_{i\alpha}$ denotes the $\alpha$ component of the spin at site i with position of the site $\textbf{r}_i$. This calculated spin structure factor is nonzero at two \textbf{q} points in momentum space. 

\section{Magnetic propagation vector} \label{AppD}

Table \ref{table:2} contains the important exchange interaction ratio between $J^{\parallel 1}$ and $J^{\parallel 3}$ for bulk NiBr$_{2}$ for pressures up to 15 GPa. As mentioned in the main text, these exchange interactions are calculated using the four-state method. From this exchange interaction ratio, we calculate the in-plane component of the magnetic propagation vector $q^{\parallel} = \frac{1}{2\pi} \arccos{\left[ \left( 1 + \sqrt{1 - 2 (J^{\parallel 1}/J^{\parallel 3}) } \right) / 4 \right]}$ which gives the magnetic unit cell size $L_{\mathrm{m.u.c.}} \sim 1/q ^{\parallel}$ \cite{Hayami_PRB_2016,Batista_RPP_2016}. With increasing pressure, the ratio $|J^{\parallel 1}$/$J^{\parallel 3}|$ decreases resulting in an increasing  $q^{\parallel}$ with pressure. Such an increase in $q^{\parallel}$ with pressure corresponds to a decreasing $L_{\mathrm{m.u.c.}}$, which means that the magnetic unit cells gets smaller with increasing pressure. 

\renewcommand{\arraystretch}{1.3}
\begin{table}[H]
\centering
\begin{tabular}{lccc}
\hline
\hline
$P$ (GPa) & $J^{\parallel 1}/J^{\parallel 3}$ & $q^{\parallel}$ & $L_{\mathrm{m.u.c.}}$ \\
\hline
0 & -2.05 & 0.098 & 10.14 \\
5 & -1.65 & 0.110 & 9.05 \\
10 & -1.45 & 0.116 & 8.59 \\
15 & -1.25 & 0.122 & 8.16 \\
\hline
\hline
\end{tabular}
\caption{Ratio of the leading intralayer exchanges $J^{\parallel 1}/J^{\parallel 3}$, in-plane component of the magnetic propagation vector magnitude $q^{\parallel}$, and magnetic unit cell length $L_{\mathrm{m.u.c.}}$ for bulk NiBr$_{2}$ at pressures ($P$) up to 15 GPa.}
\label{table:2}
\end{table}


\end{document}